\documentclass[a4paper]{article}
\usepackage{amssymb,amsmath,graphicx}
\usepackage[T1]{fontenc}
\usepackage{tipa}
\usepackage{subfig}

\title{Measurement of acoustic and anatomic changes \\
in oral and maxillofacial surgery patients} 

\makeatletter \def\name#1{\gdef\@name{#1\\}} \makeatother
\author{Daniel Aalto$^{1,2}$, Olli Aaltonen$^2$, Risto-Pekka
  Happonen$^{3,4}$, \\ P\"aivi J\"a\"asaari$^{3}$, Atle Kivel\"a$^5$, Juha Kuortti$^5$, Jean-Marc
  Luukinen$^{3,4}$, \\ Jarmo Malinen$^{5,*}$, Tiina Murtola$^5$, Riitta
  Parkkola$^6$, \\ Jani Saunavaara$^6$, Tero Soukka$^{3,4}$, Martti Vainio$^2$}

\begin{document}
\maketitle

\let\thefootnote\relax\footnote{
  $^1$Dept. of Signal Processing and Acoustics, Aalto University, Finland\\
          $^2$Institute of Behavioural Sciences (SigMe group), 
             Unversity of Helsinki,  Finland\\
               $^3$Dept. of Oral Diseases, Turku University Hospital, Finland  \\
               $^4$Dept. Oral and Maxillofacial Surgery, University of Turku, 
               Finland \\
               $^5$Dept. of Mathematics and System Analysis, Aalto University, 
               Finland \\
               $^6$Dept. of Radiology, Medical Imaging Centre of Southwest 
               Finland, University of Turku, Finland \\ 
$^*$ Corresponding author: jarmo.malinen@aalto.fi}

\begin{abstract}
We describe an arrangement for simultaneous recording of speech and
geometry of vocal tract in patients undergoing surgery involving this
area. Experimental design is considered from an articulatory phonetic
point of view. The speech and noise signals are recorded with an
acoustic-electrical arrangement. The vocal tract is simultaneously
imaged with MRI. A MATLAB-based system controls the timing of speech
recording and MR image acquisition. The speech signals are cleaned
from acoustic MRI noise by a non-linear signal processing algorithm.
Finally, a vowel data set from pilot experiments is compared with
validation data from anechoic chamber as well as with Helmholtz
resonances of the vocal tract volume.
\end{abstract}

\section{\label{IntroSec} Introduction}

\subsubsection*{Background}

Literate people in a language are taught to think that speaking is
like writing, and that a speaker produces a distinctive acoustic
pattern of energy for every distinct vowel and consonant that we
perceive, much as a typewriter produces letters. However, this
spelling out loud is not the way we speak; spelling is a far too slow
and tedious for human communication. If human speech were segmented at
the acoustic level, the task of speech perception would be simply a
matter of identifying sounds one-by-one from the speech signal,
chaining them into words, and associating these with meanings stored
in memory.

Speech, however, is not perceived, produced, or neurally programmed on
a segmental basis.  Instead, utterances are produced and perceived as
a whole. It should be emphasised that we perceive speech by virtue of
our tacit knowledge of how speech is produced. Thus, the elements of
speech are \emph{articulatory gestures}, not the sounds these phonetic
gestures produce. The gestures are the ultimate constituents of
language which must be exchanged between a speaker and a listener if
communication through language is to occur.

The human articulatory system is the only one anatomically and
neurally efficient enough to produce acrobatic manouvers of speech
organs fast, without errors, and with minimal energy. The main vocal
tract elements used in producing phonetic gestures are the lips,
tongue tip and tongue dorsum, soft palate, and the larynx. By
combining their movements in various ways meaningful linguistic units
can be built up and conveyed via sound.  Observing as well as
modelling the related biophysical features and dynamic phenomena is
far from a trivial matter even if state-of-the-art instruments and
methods (such as computational modelling based on modern medical
imaging technologies) are available. Challenging as they are, these
approaches appear quite promising for adding to our current
understanding of what happens during normal or pathological speech.

\subsubsection*{Modelling based on multi-modal data sets}

Perhaps the most important reason for using modelling and simulations
is the inherent difficulty in observing speech biophysics in test
subjects directly.

Mathematical models of human speech production have been used for
speech analysis, processing, and synthesis as well as studying speech
production acoustics for a long time; see, e.g.,
\cite{Chiba:VNS:1958,Fant:ATS:1960,Helmholtz:LT:1863}.  Many of the
earlier numerical models were based on radical simplifications of the
underlying physics and anatomic geometry, such as the Kelly--Lochbaum
model \cite{Kelly:SS:1962} and many approaches of transmission line
type; see, e.g.,
\cite{Dunn:CVR:1950,ElMasri:DTL:1998,Mullen:WPM:2006}.  Due to modern
fast and cheap computing of large scale systems, heavier numerical
acoustics models
\cite{H-L-M-P:WFWE,Lu:FES:1993,Suzuki:SVT:1993,Svancara:NMP:2004} and
computational flow mechanics models
\cite{H-U-R-V-B:AVMHGNSOVFM,S-H-R:PCFDSF3DM} have replaced earlier
approaches where higher resolution is required.  

This article has been written having in mind the wave equation (or its
resonance version, the Helmholtz equation) and Webster's horn model
for simulating the vocal tract acoustics.  These models are
well-suited for studying speech acoustics for medical purposes as well
as for basic research in speech sciences, and it depends crucially on
getting high resolution geometries of the whole speech apparatus.  It
is further expected that incorporating soft tissue and muscles into
such models, their usability would ultimately extend into studying
normal and pathological speech production from a articulatory point of
view
~\cite{Dedouch:FEM:2002,Nishimoto:ETF:2004,Svancara:NME:2006}. This
provides novel and efficient tools for planning and evaluating oral
and maxillofacial surgery and rehabilitation; see, e.g.,
\cite{Niemi:AC:2006,Vahatalo:EG:2005} for background.

Before a numerical model for speech production (such as discussed in,
e.g.,
\cite{Aalto:dt:2009,A-A-M-M-V:MLBVFVTOPI,A-A-M:LFPSGFM,H-L-M-P:WFWE})
can be used for any practical or theoretical purpose, there are always
some model parameters that need to be estimated based on measurements
from human subjects.  Such parameters, of course, include the geometry
of the vocal and the nasal tracts from the lips and nostrils to the
beginning of the trachea. To have sufficient degree of confidence in
the simulation results, any such model must have been rigorously
validated by extensively and methodically comparing simulated speech
sounds (or their characteristics) to measurements. One way of doing
the validation is by comparing the measured and the simulated formants
(related to the acoustic resonances of the vocal tract) as has been
done in \cite{A-H-K-M-P-S-V:HFAVFFCVTR} using a Helmholtz resonance
model.  In any case, the validation of the computational model depends
on recording a coupled data set: speech sound and the precise anatomy
which produces it. We emphasize that such a multi-modal data set is
quite interesting for other reasons that have little to do with
mathematical and numerical modelling \cite{P-A-A-H-M-S-V:AFVRMRISD}.

\subsubsection*{Purpose and outline of the article}

We have developed an experimental arrangement to collect a multi-modal
data set using simultaneus Magnetic Resonance Imaging (MRI) and speech
recordings as reported in~\cite{L-M-P:RSDMRI,M-P:RSDMRIII}. The
experimental arrangement includes custom hardware, software, and
experimental protocols.  During a pilot stage in June 2010, a set of
measurements were carried out on a healthy 30-year-old male subject,
confirming the feasibility of the arrangement and the high quality of
the data obtained \cite{A-H-M-P-P-S-V:RSSAMRI, Palo:WEM:2011}. These
pilot measurements also revealed a number of issues to be addressed
before tackling the main objective: obtaining a clinically relevant
data set from a large number of patients.

The purpose of this article is to describe the final experimental
arrangement including the improvements which take into account these
issues. A second pilot experiment was carried out in June 2012 on a
healthy 26-year-old male subject in order to validate the final
experimental setup. The geometric data in Fig.~\ref{JuhaGeom} as well
as the spectral envelopes of recorded vowel signals in
Figs.~\ref{SpectralEnvelopesFront}--\ref{SpectralEnvelopesBack} are
from these experiments. All patient data is excluded from this
article.

This article consists of three parts that document the main aspects of
MRI experiments during speech. In Section~\ref{PhonExpSec} the
experimental design is discussed from the phonetic and physiologic
points of view. Technical questions related to MRI and the
simultaneous speech recording are discussed in Section~\ref{TechSec}.
The acoustic instrumentation is treated only briefly, and we refer to
earlier work
\cite{A-H-M-P-P-S-V:RSSAMRI,L-M-P:RSDMRI,M-P:RSDMRIII,Palo:WEM:2011}
for details. Instead, we concentrate on the software and digital parts
of the measurement system, optimisation of the MRI sequences, and the
automated control and timing of the experiments.  The last part of
this article, Section~\ref{DSPSec}, is devoted to digital signal
processing of the recorded signals: removing acoustic MRI noise and
artefacts, extracting formants, and validating results.

\subsubsection*{Selection of the patient groups}

We describe experimental procedures that have been designed to assess
acoustic and anatomic changes in patients undergoing oral or
maxillofacial surgery which causes changes in the vocal
tract. Patients of orthognatic surgery are a particularly attractive
study group for mathematical modelling of the speech production. Not
only are these patients mostly young adults without any
significant underlying diseases, but there is a strong medical
motivation for a comparative study of their pre- and post-operative
speech as well.

Orthognatic surgery deals with the correction of abnormalities of the
facial tissues.  The underlying cause for abnormality may be present
at birth or may be acquired during the life as the result of distorted
growth.  Orthodontic treatment alone is not adequate in many cases due
to severity of the deformities. In a typical operation, the position
of either one jaw (mandible or maxilla) or both jaws is surgically
changed in relation to the skull base. The movement of the jaw
position varies usually between 5--15 mm either to anterior or
posterior direction. Such a considerable movement has a profound
effect on the shape and volume of vocal tract, resulting detectable
changes in acoustics \cite{Niemi:AC:2006}.  Although the surgery
involves mandibular and maxillary bone, changes occur also in the
position and shape of the soft tissues defining the vocal tract. Such
change is easily quantifiable using MRI.

At the time of writing of this article, seven patients of orthognatic
surgery (out of which four are female) have undergone their
pre-operative MRI examinations following the methods and protocols
described here. We expect to enroll the total of 20 patients (10
adults of both sexes) in this research.




\section{\label{PhonExpSec}Experimental arrangement}

Generally speaking, the experimental setting is similar to the setting
in which the pilot arrangement was
tested~\cite{A-H-M-P-P-S-V:RSSAMRI,Palo:WEM:2011} but with numerous
improvements. They are related to instructing and cueing the patient,
the role of the experimenter, and the automated control and timing of
MR imaging.

The creation of the original pilot data reported in
\cite{A-H-M-P-P-S-V:RSSAMRI} required $3$ -- $4$ people working
simultaneously in the MRI control room. The improved arrangement
described in this article requires only two people: one for MR imaging
and the other for running the integrated experimental control system
and sound recording. Moreover, it is now possible to produce over $60$
takes during a session of $1$ h which is about three times as fast
data a collection rate as can be attained using a non-automated
system. The streamlining of all procedures is vital because laboratory
downtime and cost must be minimised when gathering a large data
set. Also patient comfort and performance are compromised by overly
long MRI sessions.

\subsection{Phonetic material}

The speech materials have been chosen to provide a phonetically rich
data set of Finnish speech sounds. The chosen MRI sequences require up
to 11.6 s of continuous articulation in a stationary position. We use
the Finnish speech sounds for which this is possible: vowels
\textipa{[\textscripta , e, i, o, u, y, \ae , \oe]}, nasals
\textipa{[m, n]}, and the approximant \textipa{[l]}. A long phonation
is possible also for, e.g., \textipa{[j, s, N]} but these have been
excluded because of unpleasantness in supine production
(\textipa{[N]}) and turbulences in the vocal tract (\textipa{[j,~s]}).

Patients are instructed to produce each of the sounds at a sustained
fundamental frequency ($f_0$). We use two different $f_0$ levels (104
and 130~Hz for men, 168 and 210~Hz for women) for the sounds
\textipa{[\textscripta]} and \textipa{[i]} to obtain the vocal tract
geometry with different larynx positions. The rest of the sounds are
produced at the lower $f_0$ only. The $f_0$ levels have been matched
with the acoustic MRI noise frequency profile to avoid interference.

In a sustained phonation, the long exhalation causes contraction in
the thorax and hence a change in the shape of vocal organs. The
stationary 3D imaging sequence which is used to obtain the vocal tract
geometry provides no information on this adaptation process, so
additional dynamic 2D imaging on the mid-sagittal section for the
sounds \textipa{[\textscripta , i, u, n, l]} is used to monitor
articulatory stability.
 
Speech context data is also acquired by having the patient repeat 12
sentences which have been selected from a phonetically balanced
sentence set~\cite{V-S-J-J-M:DSITBMSRTNEF}. A few syllable
repetition measurements (``tattatta'', ``kikkikki'', etc.) are
done, too. These continuous speech samples are imaged using the same
dynamic 2D sequence which is used for checking articulatory stability.

\begin{figure}[t]
\centerline{\includegraphics[width=80mm]{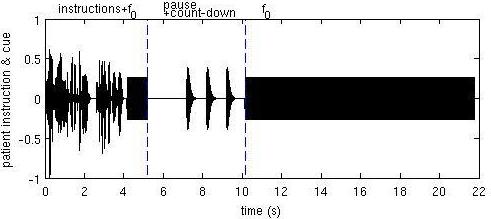}}
\caption{{\it Patient instruction and cue signal structure.}}  
\label{cue}
\end{figure}

An instruction and cue signal is used to guide the patient through
each measurement. The signal consists of three parts as shown in
Fig.~\ref{cue}: (i) recorded instructions specifying the task with a
sample of the desired $f_0$, (ii) a 2 s pause and three count-down
beeps one second apart, and (iii) continuous $f_0$ for 11.6 s. In case
of speech context experiments, the recorded instructions specify the
sentence to be repeated and $f_0$ is left empty in both parts (i) and
(iii).  Audibility of the $f_0$ cues over MR imaging noise is achieved
by using a sawtooth waveform.

\subsection{Setting for experiments}

The patient lies supine inside the MRI machine with a
sound collector placed on the Head Coil in front of the patient's
mouth.  The patient can communicate with the control room through the
sound collector and the earphones of the MRI machine. The patient can
also hear his own (de-noised) voice through the headphones with a
delay of approximately 90 ms.

The patients familiarise themselves with the tasks and the phonetic
materials before the beginning of a measurement session. They also
practice the tasks under the supervision and are given feedback on
their performance.  At the start of a measurement, the experimenter
selects the phonetic task. The patient then hears the recorded
instructions. The instructions and the following pause and count-down
beeps give the patient time to swallow, exhale, and inhale before
phonation is started after the count-down beeps. The patient hears the
target $f_0$ in the earphones added to his own (de-noised) voice
throughout the phonation.

MR imaging for static 3D and dynamic stability check sequences is
started 2~s after the start of phonation and finishes approximately
500~ms before the end of phonation. Thus ``pure samples'' of
stabilised utterance are available before and after the imaging
sequence. Two 200~ms breaks are inserted into the MRI sequences.  The
duration of these breaks has been determined based on the half-time of
the imaging noise in the MRI room, which was measured to be
approximately 20~ms. Sentence and syllable imaging sequences start
simultaneously with phonation and end after 3.2~s.

The experimenter listens to the speech sound throughout the
experiment, allowing unsuccessful utterances to be detected
immediately. At the end of the experiment, the experimenter writes
comments and observations into a metadata file. The recorded sound
pressure levels are also inspected. Unsuccessful measurements are
repeated, at the experimenter's discretion, either immediately or
later in the measurement set.

\section{\label{TechSec}Simultaneous MRI and speech recording}

The MRI room presents a challenging environment for sound recording
due to acoustic noise and interference to electronics from the MRI
machine. For safety and image quality reasons, use of metal is
restricted inside the MRI room and prohibited near the MRI
machine. Our approach is to use passive acoustic instruments for
collecting the sound samples and transmitting them to a safe distance
from the MRI machine. Alternative solutions are {\rm (i)} using an
optical microphone inside the MRI machine \cite{Opto:2009}, {\rm (ii)}
recording by conventional directional microphones sufficiently far
away from the MRI machine that has an open construction
\cite{P-H-H:TMMNRRSDPMRID,P-P-F:ASPANPDMRI}, and {\rm (iii)} using the
internal microphone of the MRI machine \cite{Svancara:NME:2006}.

\subsection{\label{SpeechRecSec}Speech recording}

\begin{figure}[t]
  \centering
  \subfloat[]{
    \label{sc}
    \includegraphics[height=35mm]{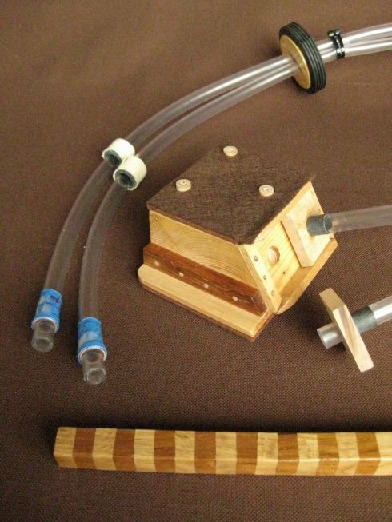}
  }
  \hspace{0.2cm}
  \subfloat[]{
    \label{faraday}
    \includegraphics[height=35mm]{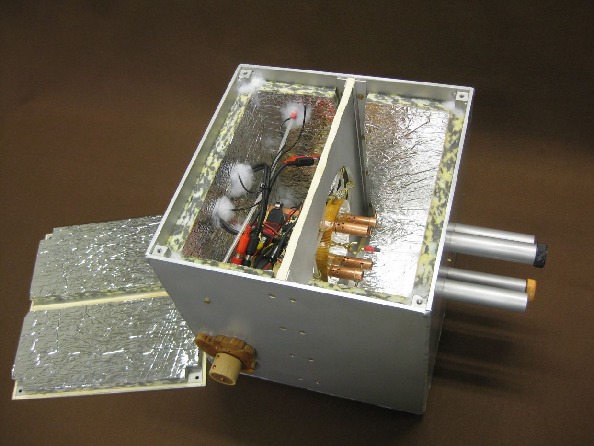}
  }
  \caption{ (a) The sound collector with one of the two audio wave
      guides attached. (b) The microphone array inside the Faraday
      cage.}

\end{figure}

We use instrumentation specially developed for speech recording during
MRI~\cite{L-M-P:RSDMRI,M-P:RSDMRIII}: A two-channel sound collector
(Fig.~\ref{sc}) samples the speech and primary noise signals in a
dipole configuration. The sound signals are coupled to a microphone
array inside a sound-proof Faraday cage (Fig.~\ref{faraday}) by
acoustic waveguides of length 3.00~m (the ends of which can be seen in
Fig.~\ref{sc}). Two additional ``ambient noise'' samples are
collected: one from the microphone array inside the Faraday cage and
another from inside the MRI room using a directional electret
microphone near the patient's feet, pointing towards the patient's
head and the MRI coil.

The four signals are coupled from the microphones to a custom RF-proof
amplifier that is situated in the measurement server rack (shown in
Fig.~\ref{server}) outside the MRI room. The amplifier contains
additional circuitry (i.e., a long-tailed pair with a constant emitter
current source) for optimal, real time analogue subtraction of the
primary noise channel from the speech channel. This is intended to
produce the de-noised signal played back to the patient, and it is
used for audio signal quality observation in the MRI control room as
well. The final, high-quality de-noised signal is not produced this
way but from digitised component signals by the algorithm discussed in
Section~\ref{DSPSec}. We remark that the hardware appears to be able
to transmit good signal at least up to 10~kHz but we use only the
phonetically relevant frequency range below 4.5~kHz where the
measured frequency response is given in Fig.~\ref{freqres}.

Audio signals are converted between analogue and digital forms using a
M-Audio Delta 1010 PCI Audio Interface.  A measurement server is used
which has an Intel Core i7-860 processor clocked at 2.80GHz, and is
equipped with 4Gb RAM and a SSD drive for fast booting. For immediate
internal data backup, three additional 1.5TB discs are set up in RAID1
configuration by a HighPoint RocketRaid 2302 controller. The whole
setup is powered by an APC Smart-UPS SC 450VA, and it is installed to
a portable 10U rack as shown in Fig.~\ref{server}. All user access to
the server is done with laptops (in fact, MacBooks) running X11
servers, either via 1GB LAN or a wireless access point.

\subsection{Magnetic resonance imaging}

\begin{figure}[t]
  \centering
  \subfloat[]{
\label{JuhaSurface}
    \includegraphics[height=45mm]{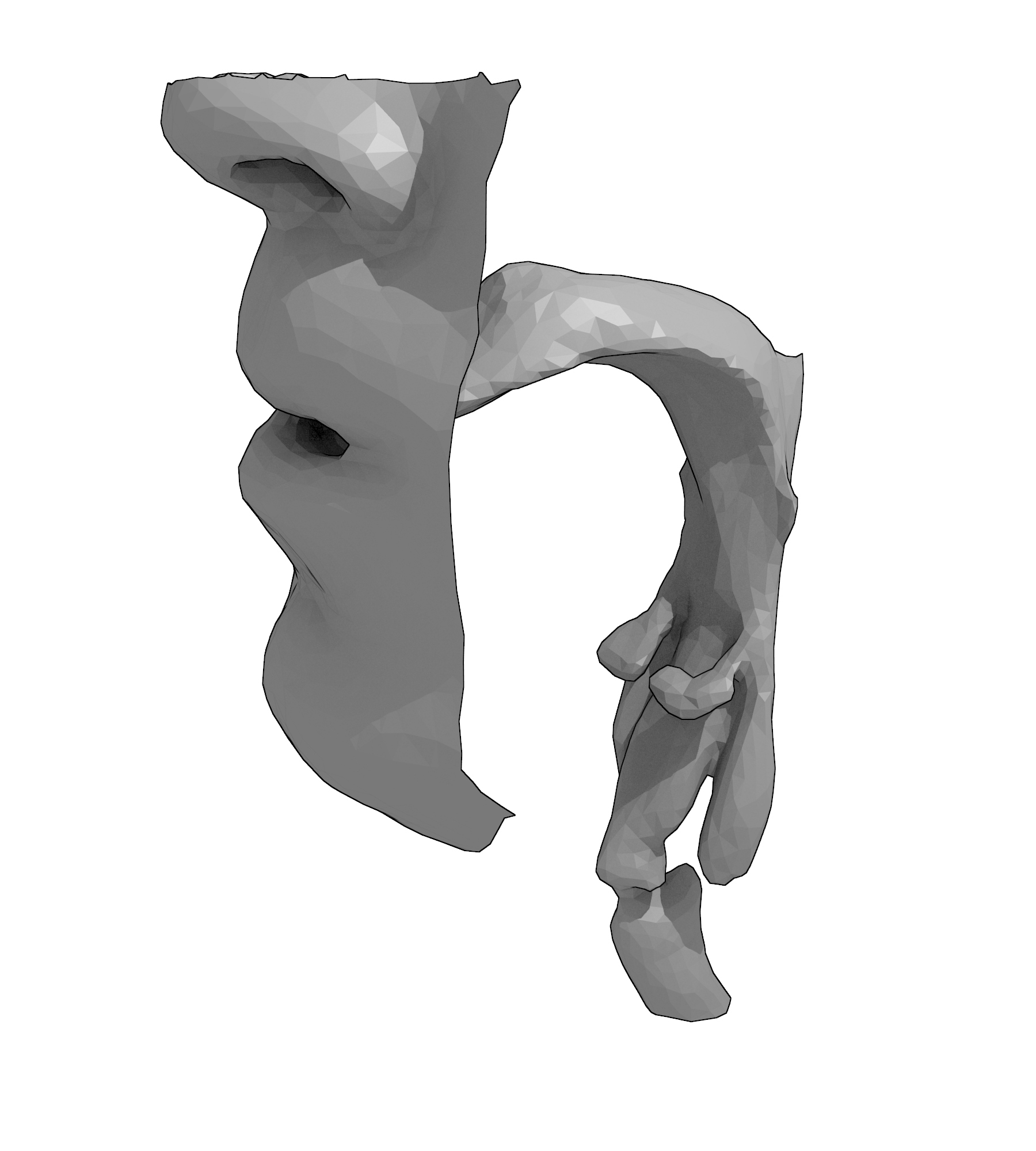}
  }
\vspace{5mm}
  \subfloat[]{
\label{JuhaArea}
    \includegraphics[height=40mm]{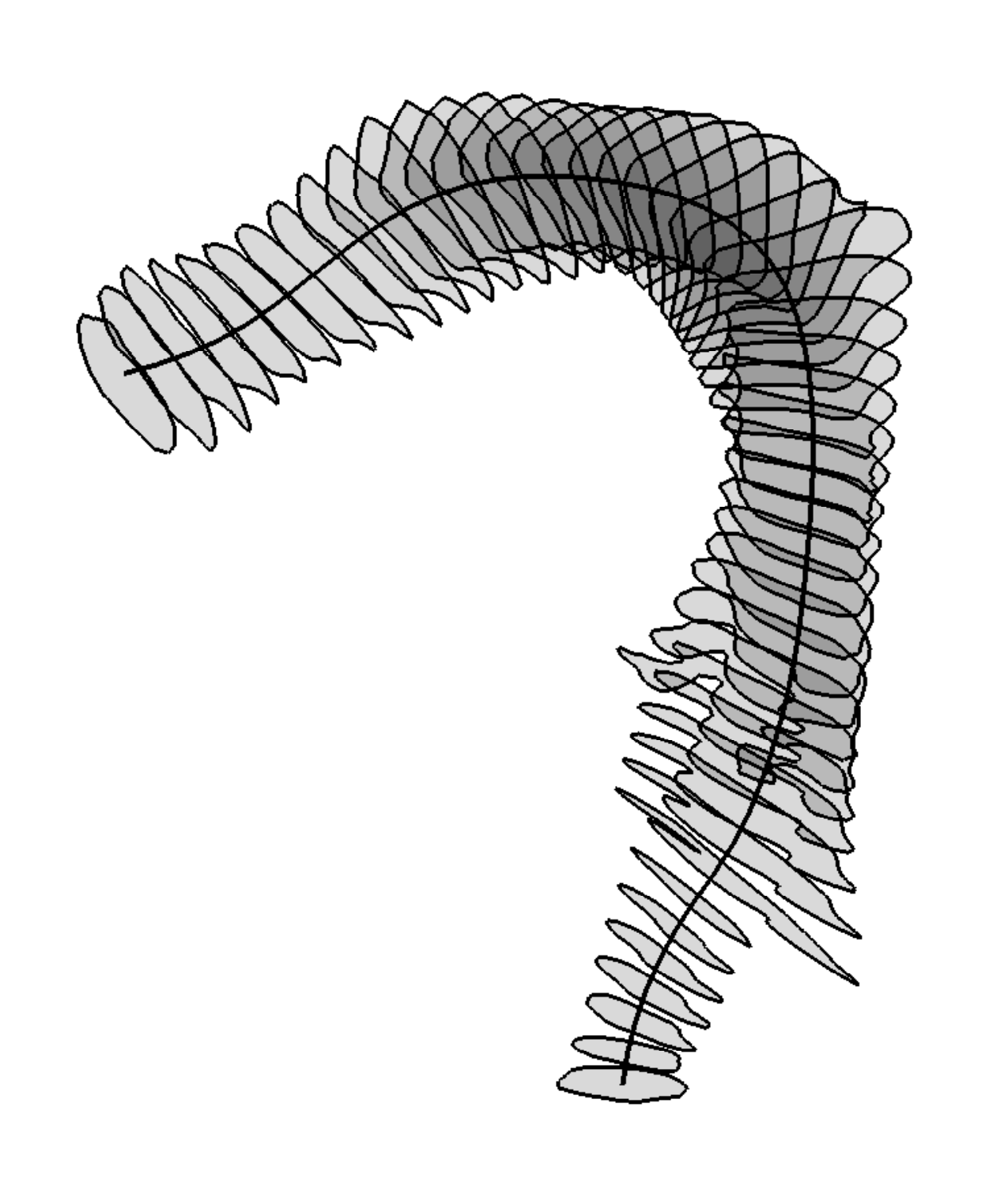}
  }
  \caption{ (a) The surface model of the tissue-air interface of a
    male vocal tract while pronouncing [\textipa{\oe}].  (b) The
    centreline and intersection areas extracted from the same
    geometry. }
\label{JuhaGeom}
\end{figure}

Measurements are performed on a Siemens Magnetom Avanto 1.5T scanner
(Siemens Medical Solutions, Erlangen, Germany). Maximum gradient field
strength of the system is 33~mT/m (x,y,z directions) and the maximum
slew rate is 125~T/m/s. A 12-element Head Matrix Coil and a 4-element
Neck Matrix Coil are used to cover the anatomy of interest. The coil
configuration allows the use of Generalized Auto-calibrating Partially
Parallel Acquisition (GRAPPA) technique to accelerate
acquisition. This technique is applied in all the scans using
acceleration factor 2.

3D VIBE (Volumetric Interpolated Breath-hold Examination) MRI
sequence~\cite{Rofsky:VIBE:1999} is used as it allows for the rapid
static 3D acquisition required for the experiments. Sequence
parameters have been optimized in order to minimize the acquisition
time. The following parameters allow imaging with 1.8~mm isotropic
voxels in 7.8~s: Time of repetition (TR) is 3.63~ms, echo time (TE)
1.19~ms, flip angle (FA) $6^\circ$, receiver bandwidth (BW)
600~Hz/pixel, FOV 230~mm, matrix 128x128, number of slices 44 and the
slab thickness of 79.2~mm.

Dynamic MRI scans are performed using segmented ultrafast spoiled
gradient echo sequence (TurboFLASH) where TR and TE have been
minimized. Single sagittal plane is imaged 
using parameters TR 178~ms, TE 1.4~ms, FA 6$^\circ$, BW
651~Hz/pixel, FOV 230~mm, matrix 120x160, and slice thickness 10~mm.

Siemens Magnetom Avanto 1.5T units have inputs that accept external
syncronisation signals for timing the MRI sequences.  We use external
triggering in all three different types of experiments, and the
triggering signal is always a train of 12 ms TTL level~1 pulses
separated by TTL level~0 of variable duration.  The pulse train is
generated with a custom-made device which converts 1 kHz analogue sine
signal from the sound system to the logic pulses. The details of
triggering are given in Table~\ref{TriggeringParameters}.

External triggering with the additional pauses increases the 3D
imaging time to 9.1 s.  Post-processing of the MR images and the
resolution of the obtained vocal tract geometries are discussed
in~\cite{A-H-H-K-M-S-R:ASEMRIDMHVT,A-H-K-M-P-S-V:HFAVFFCVTR,A-M-V-S-P:EMAEMRID}.

\subsubsection*{Visibility of teeth} 

Teeth are not visible in MRI but they are an important acoustic
element of the vocal tract. Hence, it is necessary to add teeth
geometry into the soft tissue geometry obtained from the MR images
during post-processing. Optical scans of teeth or digitalised dental
casts can be readily obtained from the patients but the alignment of
the two geometries is a non-trivial problem. Markers containing
vegetable oil, attached to the surface of the teeth, appear to be a
practical approach that produces sufficient MRI visibility; see
also~\cite{Ericsdotter:AAR:2005}. Further work is still required to
get a solution for alignment that does not require extensive manual
work.

\begin{table}[t]
  \centerline{
    \begin{tabular}{|c|c|c|c|c|c|}
      \hline  
          & 3D static & 2D stability  & 2D sentence/syllable   \\
      \hline  
      Pulse separation      & 240~ms & 140~ms & 150~ms    \\
      Number of slices      & 35  & 69  & 20     \\
      Pause after slice     & 12 and 24 & 23 and 43 & no pause \\
      \hline
  \end{tabular}}
  \caption{\label{TriggeringParameters} External triggering parameters
    used in MRI scans.}
\end{table}

\subsection{Control of measurements}

\begin{figure}[t]
  \centering \subfloat[]{
    \label{server}
    \includegraphics[height=35mm]{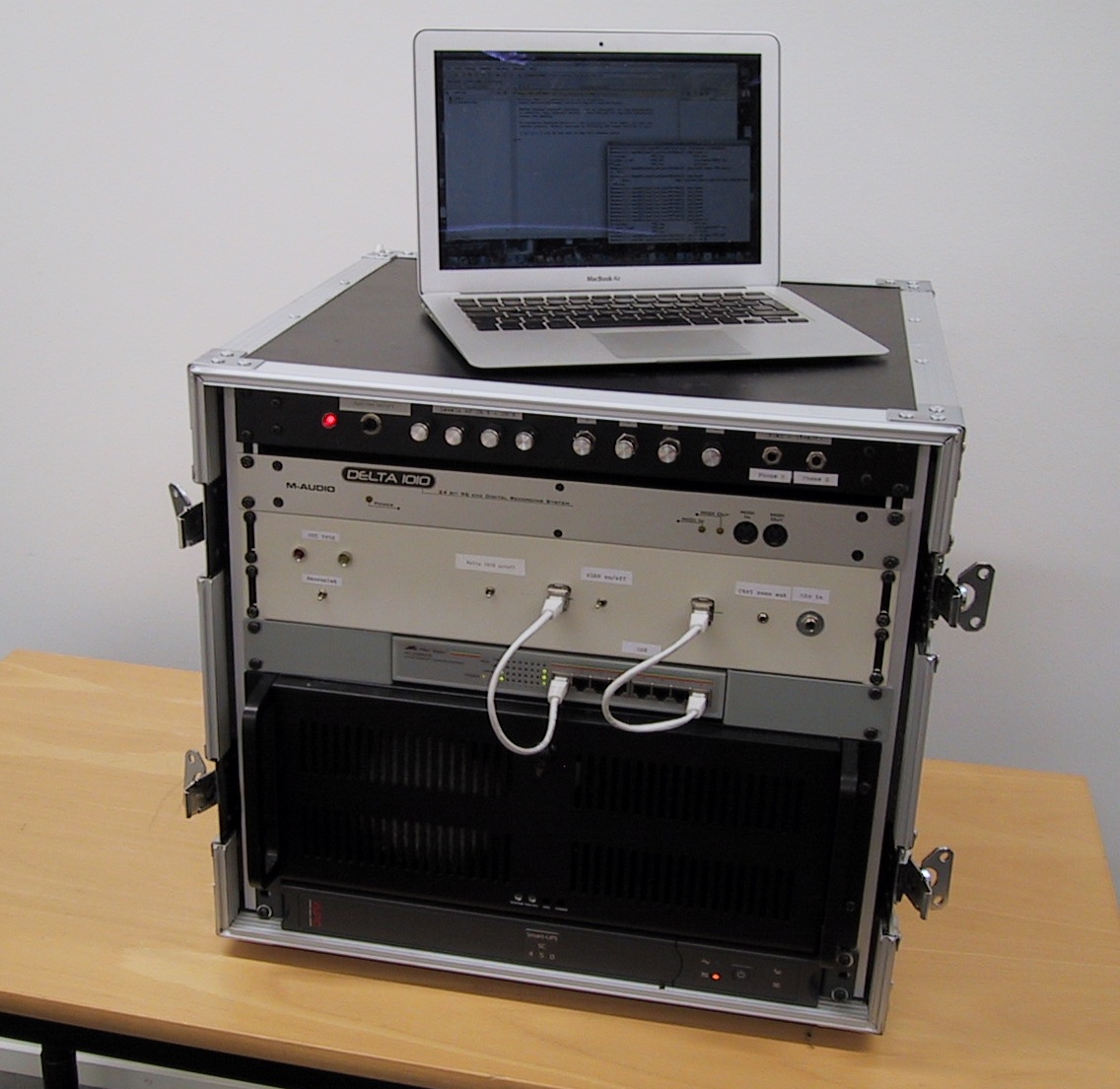}
  }
  \subfloat[]{
    \label{freqres}
\vspace{1cm}
    \includegraphics[height=35mm]{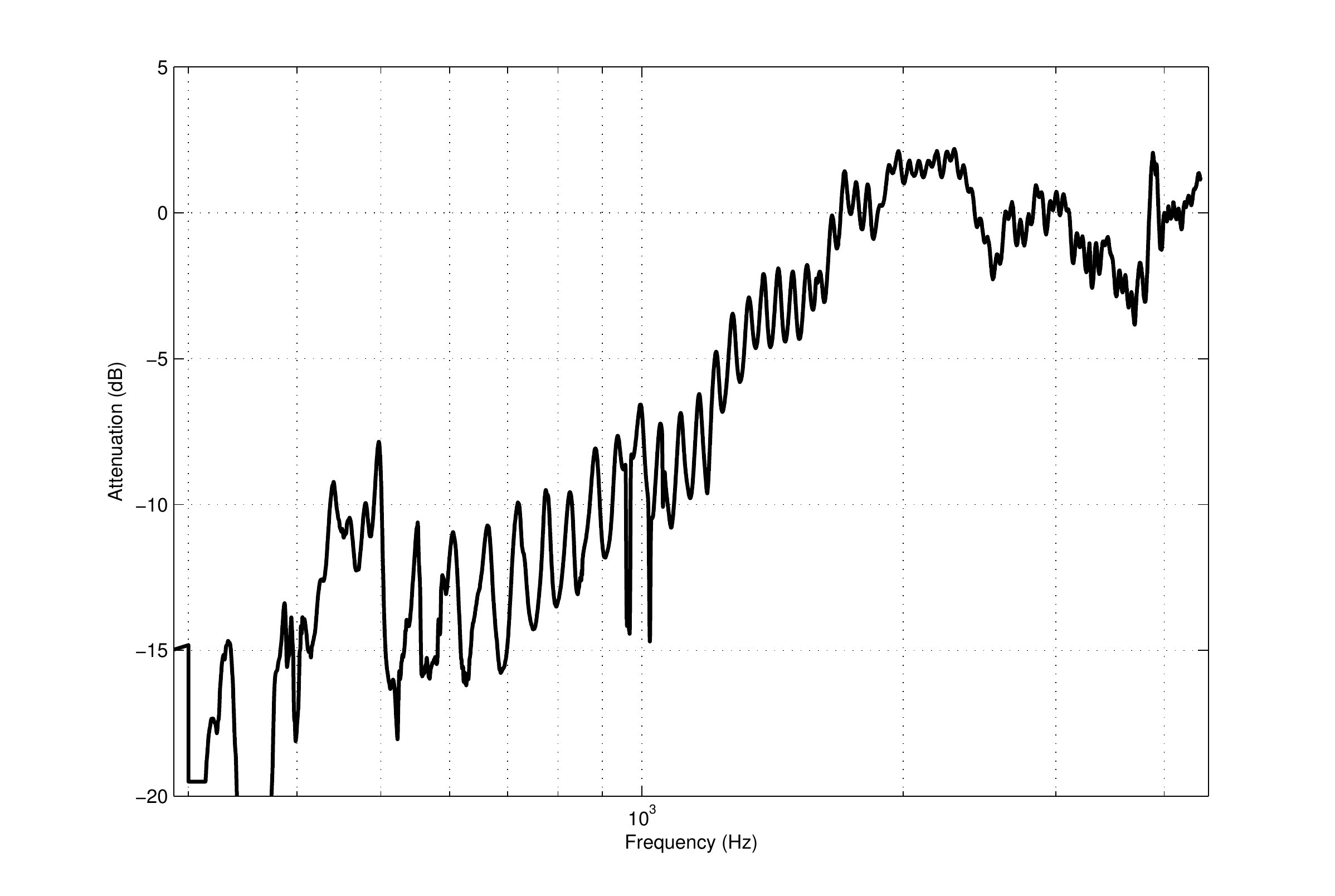}
  } 
\caption{ (a) The measurement server with the RF-proof four-channel
  amplifier, M-Audio Delta 1010 Audio Interface, and networking
  facilities. The laptop is used for remote access.  (b) The measured
  frequency response of the whole audio signal path from the sound
  collector to the amplifier input. The non-flat frequency response is
  compensated by DSP in post-processing stage. }
\end{figure}

Measurements are controlled with a custom code in MATLAB 7.11.0.584
(R2010b) running on the portable server with operating system Ubuntu
10.04 LTS on Linux 2.6.32.38 kernel (Fig.~\ref{server}).  Access from
MATLAB to the Audio Interface is arranged through Playrec (a MATLAB
utility,~\cite{Playrec}), QjackCtl JACK Audio Connection Kit
(v.~0.3.4), and JackEQ 0.4.1.

The custom code computes the input signal to the MRI triggering
device, reads the patient instruction and cue audio file, and
assembles the two signals into a playback matrix. Recording is started
simultaneously with playback and carried on for equal number of
samples. In addition to the speech and the three noise signals,
recording also includes the analogically de-noised signal and the
patient instruction signal.

The audio configuration causes delays in the signals. MRI noise is
recorded with a delay of approximately 60 ms (MRI machine delays
excluded) relative to the onset of recording. This is accounted for by
the method of locating the ``pure samples''. Patient speech is
recorded with a delay of approximately 90 ms (patient reaction time
excluded), and patients also hear their own voices with this same
delay. These patient speech delays may have two effects. First, the
duration of the last pure sample is reduced from 500 ms to 410 ms,
which makes no significant difference from the point of view of data
analysis. And second, the echo effect may disturb the patient during
sentence repetition in which case the speech feedback may be turned
off or its volume reduced independent of the cue signal.

The control code automatically saves the recorded sounds as a
six-channel Waveform Audio File. A separate file containing metadata
is also saved automatically. The metadata file contains all
experimental parameters, including task specification, and the
locations of the pure samples in the sound file.

The control system requires user input for three tasks. First, the
experimenter selects the next phonetic task (target sound or sentence
and $f_0$) and MR imaging sequence. Second, comments and observations
may, if necessary, be written about each measurement separately. They
are saved automatically in the metadata file in JSON format. And
third, patient headphone volume and recorded sound pressure levels may
be adjusted manually based on feedback from the patient and
rudimentary post-experimental sound data checks. The sound data checks
consist of histograms of recorded signal levels,
and they are displayed to the experimenter automatically at the end of
each measurement.  This allows detection and correction of settings
for which the recorded signal levels, which vary for different speech
sounds, are outside the optimal range.

A single measurement takes on average 30-40 s, including task
selection by the experimenter and writing additional information and
observations in the metadata file. At the time of writing of this
article, seven patients of orthognatic surgery have taken part in the
experiments, and the times spent inside the MRI machine were between
50--95 min.  When running the experiment at a comfortable pace for the
patient, between 93 and 107 MRI scans were produced in a single
session.

\section{\label{DSPSec}Post-processing of speech signals}

\subsection{\label{AlgorithmSec}Adaptive noise reduction}

\begin{figure}[t]
  \centering
  \subfloat[]{
\label{flow}
\includegraphics[width=65mm]{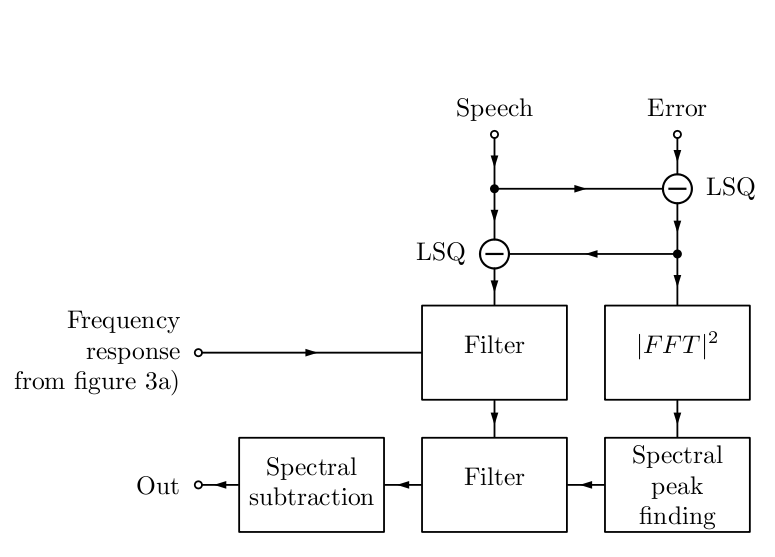}
}
  \centering 
  \subfloat[]{
\label{SpectralPeaks}
\includegraphics[width=50mm]{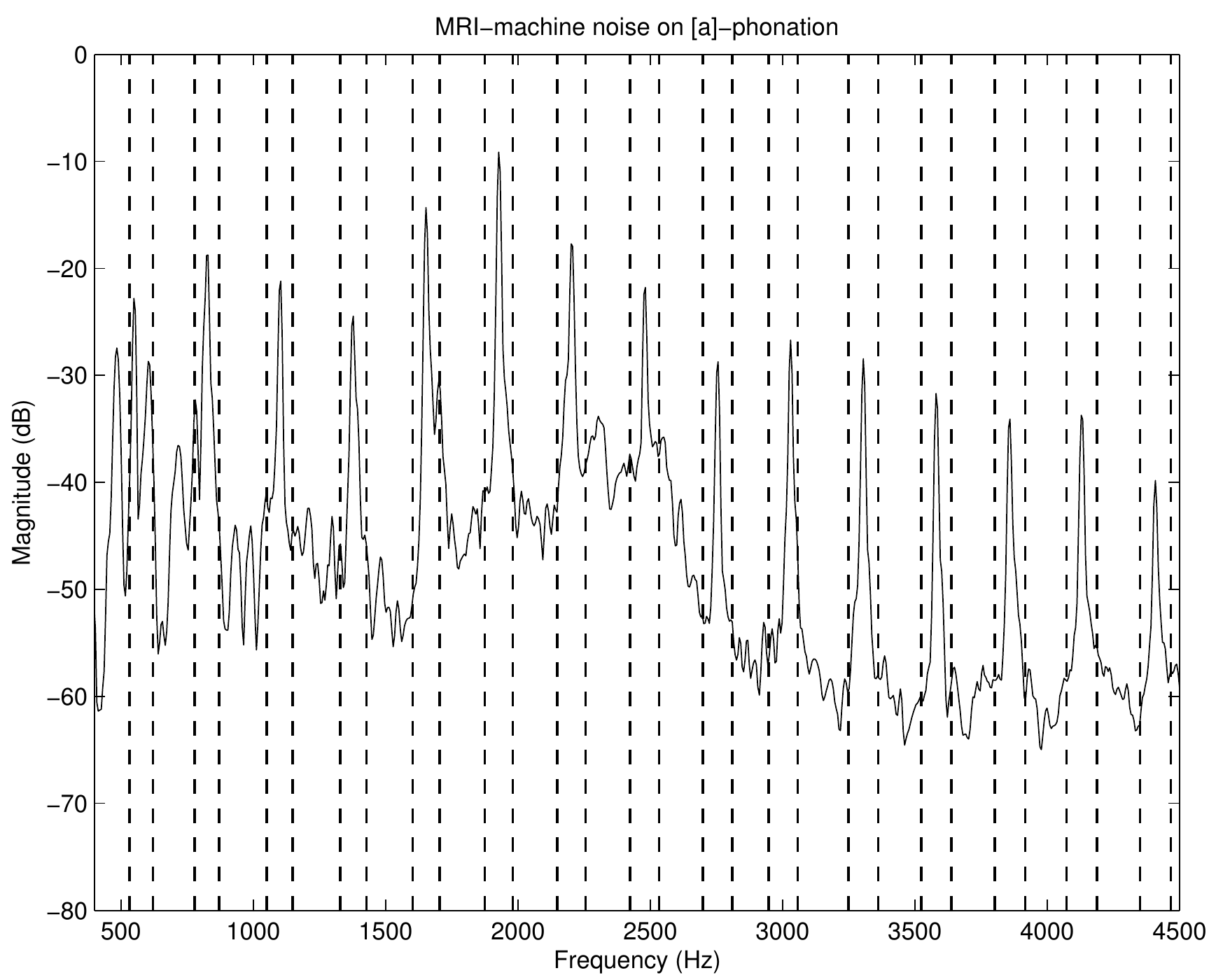}
}
\caption{ (a) Block diagram of the noise reduction algorithm. (b)
  The detected spectral noise peaks. Note the regular harmonic
  structure.}
\end{figure}
 
As explained in Section~\ref{SpeechRecSec}, two sound channels are
acoustically transmitted from near the test subject inside the MRI
machine. One of the channels provides the speech sample $s(t)$ (which
is contaminated by acoustic MRI noise), and the other is reserved for
the acoustic MRI noise sample $n(t)$ (which, in turn, is contaminated
by speech). The analogically produced weighted difference of these
signals is fed back to subject's earphones during the experiment in
almost real time. Both the signals $s(t)$ and $n(t)$ are also recorded
separately, so that more refined numerical post-processing can be
performed later.

Because of the multi-path propagation of the noise in particular
around the MRI coil surfaces, the recorded noise sample is a weighted
sum of more simple signals with distributed delays. As a further
complication, the chassis of the MRI machine acts as a spatially
discributed acoustic source, and its dimensions are large compared to
wavelengths in air at frequencies of interest. Hence, some residual
higher frequency noise will remain after an optimised subtraction of
the noise $n(t)$ from the contaminated speech signal $s(t)$. To reduce
this residual noise, \emph{adaptive spectral filtering} is used. The
approach is based on the observation that the typical noise spectrum
of a MRI machine consists of narrow and high peaks with significant
harmonic overtones. Adaptivity is desirable because the peak positions
depend on the MRI sequence used, and they are not invariant of time
even within a single MRI sequence.

The noise reduction algorithm is outlined in Fig.~\ref{flow}, and it
consists of the following Steps~\ref{LSQ}--\ref{SpecSub} that have
been realised as MATLAB code:
\begin{enumerate}
\item \label{LSQ} {\bf LSQ:} Linear, least squares optimal subtraction
  of the noise from speech as detailed below. This reproduces roughly
  the same quality of speech signal that was produced analogically
  during the experiment for patient's earphones in real time.
\item \label{FreqResComp} {\bf Frequency response compensation:} Compensation for the
  measured non-flat frequency response of the measurement system, show
  in Fig.~\ref{freqres}.
\item {\bf Noise peak detection:} The noise power spectrum is computed
  by FFT, and the most prominent spectral peaks of noise are detected.
\item {\bf Harmonic structure completion:} The set of noise peaks is
  completed by its expected harmonic structure to ensure that most of
  the noise peaks have been found; see Fig.~\ref{SpectralPeaks}.
\item \label{Chebyshev1} {\bf Chebyshev peak filtering:} Each of the
  noise peaks defines a centre frequency of a corresponding stop
  band. The width of the stop band is a function of the centre
  frequency given by Eq.~\eqref{StopBandWidth}. The corresponding
  frequencies are attenuated from the de-noised speech signal (that
  has been produced in Step~\ref{FreqResComp} above) by Chebyshev
  filters of order $20$ at these stop bands.
\item \label{Chebyshev2} {\bf Low pass filtering:} The resulting
  signal is low pass filtered by a Chebyshev filter of order $20$ and
  cut-off frequency $10$ kHz.
\item \label{SpecSub} {\bf Spectral subtraction:} A sample of the
  acoustic background of the MRI room (without patient speech and the
  noise during the MRI sequence) is extracted from the beginning of
  the speech recording. Finally, the averaged spectrum of this is
  subtracted from the speech signal in frequency plane using FFT and
  inverse FFT; see \cite{SB:SANSUSS}.
\end{enumerate} 
The optimal linear subtraction in Step~\ref{LSQ} is carried out by
producing de-noised signals $\tilde s(t)$, $\tilde n(t)$ from the
original signals $s(t)$ and $n(t)$ according to
\begin{equation} \label{OptimalSubtraction}
   \tilde n  = n - \frac{\left < n, s \right >}{||n|| \cdot ||s|| } s
   \quad \text{ and } \quad \tilde s  = s - \frac{ \left < s, \tilde n
     \right >}{ ||s|| \cdot ||\tilde n||} \tilde n 
\end{equation}
where $\left < n, s \right > = \int{n(t) s(t) \, dt}$ and $|| s || ^2
= \left < s, s \right >$.  The bandwidths in Step~\ref{Chebyshev1} are
given as a function of the centre frequency $f$ by
\begin{equation} \label{StopBandWidth}
  w(f) = C \ln{f} \quad \text{ where } \quad  w(550 \text{ Hz}) = 50 \text{ Hz}.
\end{equation}
The numerical parameters values (i.e., the bandwidth parameter $C$,
the filter order, and the cut-off frequency) have been determined by
trial and error to get audibly good separation of speech and noise in
prolonged vowel samples. In particular, choosing the bandwidth
parameter $C$ for Step~\ref{Chebyshev1} is crucial for the
outcome. The cut-off frequency of 10 kHz in Step~\ref{Chebyshev2} is
chosen well above the phonetically relevant part of the frequency
range that extends upto 4.5 kHz corresponding to Fig.~\ref{freqres}.

The algorithm produces de-noised speech signals where the S/N ratio is
audibly much improved compared to the mere optimal subtraction as
defined in Eq.~\eqref{OptimalSubtraction}. Each speech sample contains
2 s of undisturbed speech before acoustic MRI noise starts, and
comparing the amplitude of the speech channel signal just before and
right after the noise onset, we can get an estimate for the S/N ratio
(assuming that the speech amplitude remains reasonably constant at the
MRI noise onset, and that speech and noise are
uncorrelated). Moreover, the S/N ratio depends on the vowel because
the emitted acoustic power tends to be larger for vowels with larger
mouth opening area. As a rule of thumb, we obtain cleaned-up vowel
signals whose S/N ratio lies between 1.9~dB and 3.6~dB, the average
being 2.8~dB. The Chebyshev filtering in Step~\ref{Chebyshev1} creates
a somewhat ``robotic'' tone to speech signals but we have not carried
out perceptual evaluation of the de-noised signals as was done in
\cite{P-A-A-H-M-S-V:AFVRMRISD}.

The subtraction of noise with a ``spiky'' power spectrum from, e.g.,
speech is a classical problem in audio signal processing. The
non-linear \emph{cepstral transform} is a popular procedure, and it
has been used successfully in \cite{P-H-H:TMMNRRSDPMRID} for MRI noise
cancellation. This algorithm is based on computing the logarithm of
the power spectrum (in order to compress all high spectral peaks
``softly'' and non-adaptively), returning to time domain by FFT, and
reconstructing the phase information from the original signal. The
cepstral transform does not take into account the harmonic structure
of noise at all. The multi-path propagation of noise would seem to
invite an approach based on \emph{deconvolution}. However, an accurate
estimation of the convolution kernel (i.e., the delays and the weights
in multi-path propagation) does not seem to be feasible even though
the autocorrelation of the noise signal is easy to compute.

\subsection{\label{FormantSec}Extracting power spectra and spectral envelopes} 

Formants are the main information bearing component of vowel
sounds. They can be understood as acoustic energy concentrations
around discrete frequencies in the power spectrum of the speech
signal.  The measured formant frequencies $F_1, F_2, \ldots$ are
related to the acoustic resonance frequencies $R_1, R_2, \ldots$ of
the vocal tract. In contrast to harmonic overtones of the fundamental
frequency $f_0$ of the glottal excitation, the formants have a much
wider bandwidth. Thus, the extraction of formants from speech can be
carried out by a frequency domain smoothing process that downplays the
narrow bandwidth harmonics of $f_0$.

Perhaps the most popular formant extraction tool is Linear Predictive
Coding (LPC); see, e.g., \cite{JM:LPATR,Makhoul:SLP:1975}.  LPC is
mathematically equivalent to fitting a low-order rational function
$R(s)$ to the power spectrum function defined on the imaginary axis,
and the pole positions of $R(s)$ give the estimated formant
values. Hence, plotting the values of $|R(i \omega)|$ for real
$\omega$ yields \emph{LPC envelopes} whose peaks indicate the formant
frequencies $F_1, F_2, \ldots$.

\begin{figure}[t]
\center
\includegraphics[width=55mm,height=40mm]{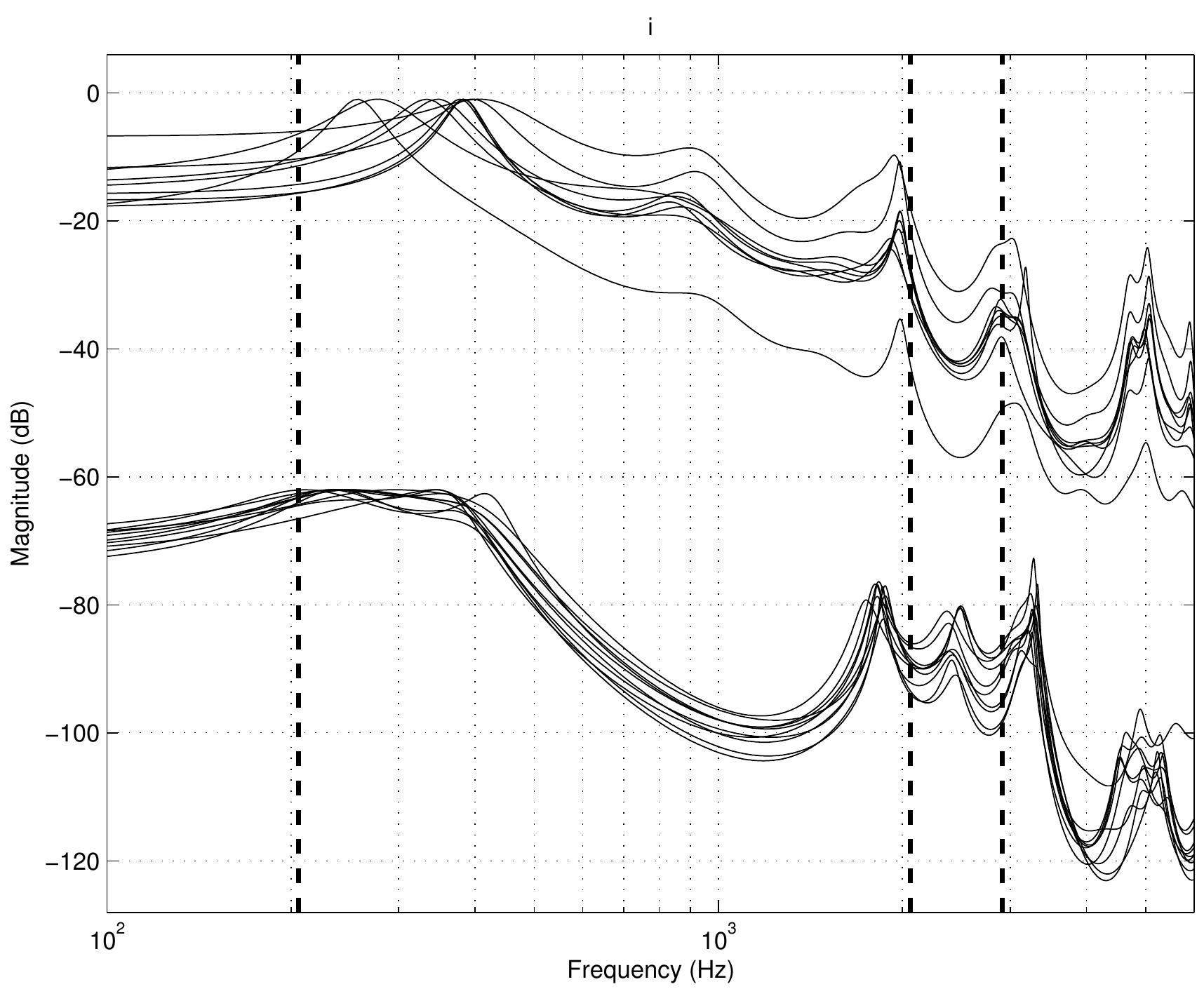}
\includegraphics[width=55mm,height=40mm]{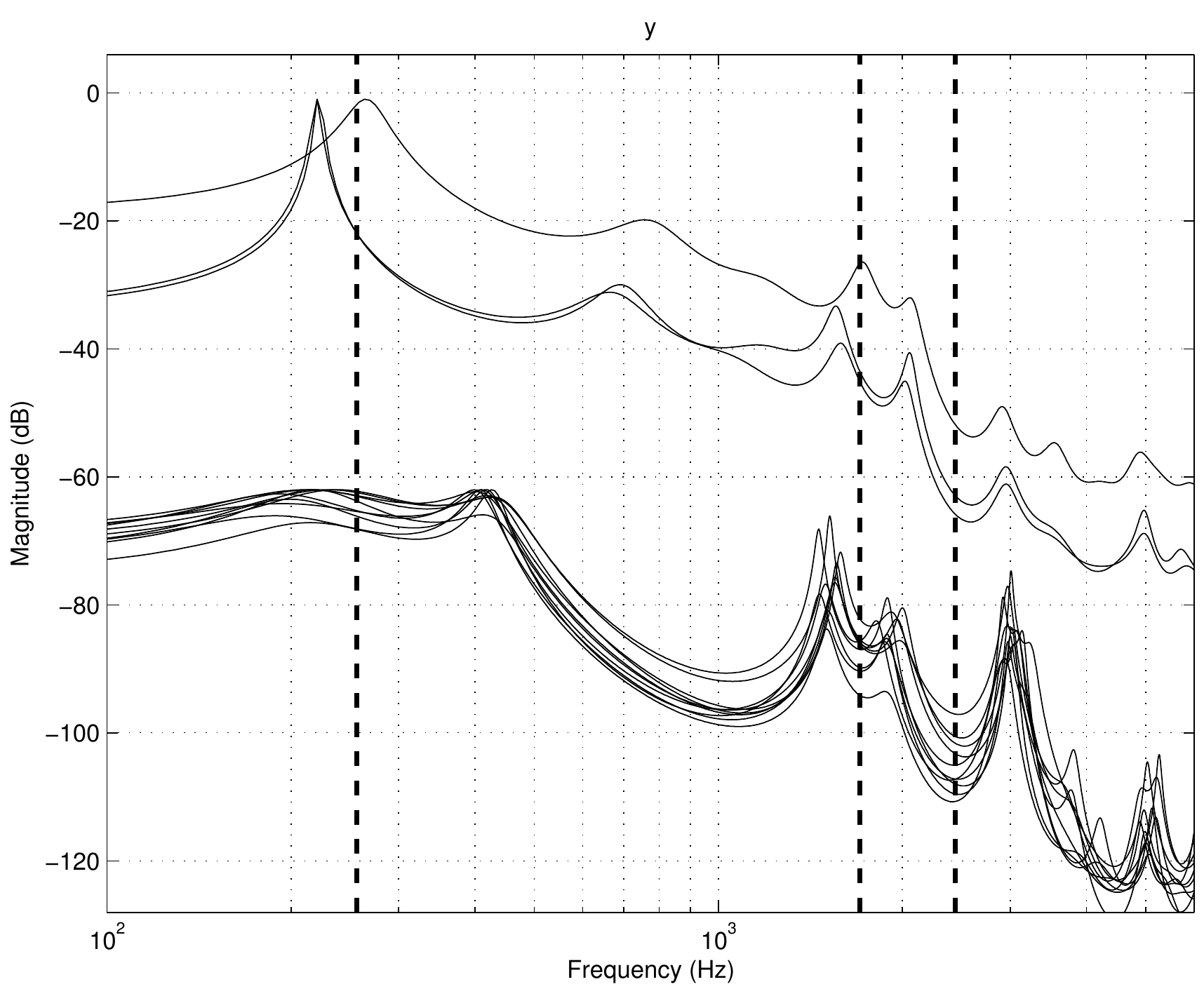}
\center
\includegraphics[width=55mm,height=40mm]{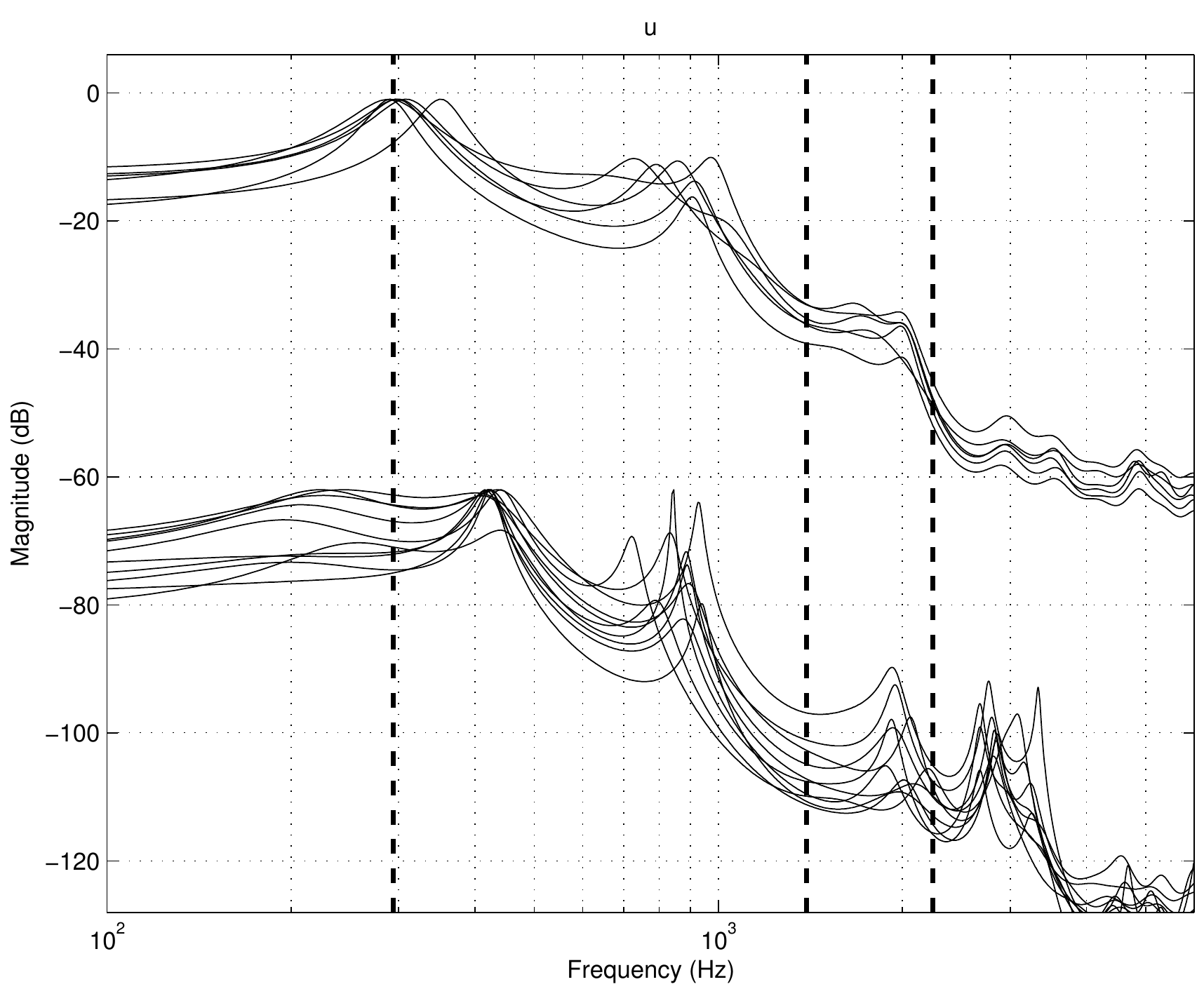}
\includegraphics[width=55mm,height=40mm]{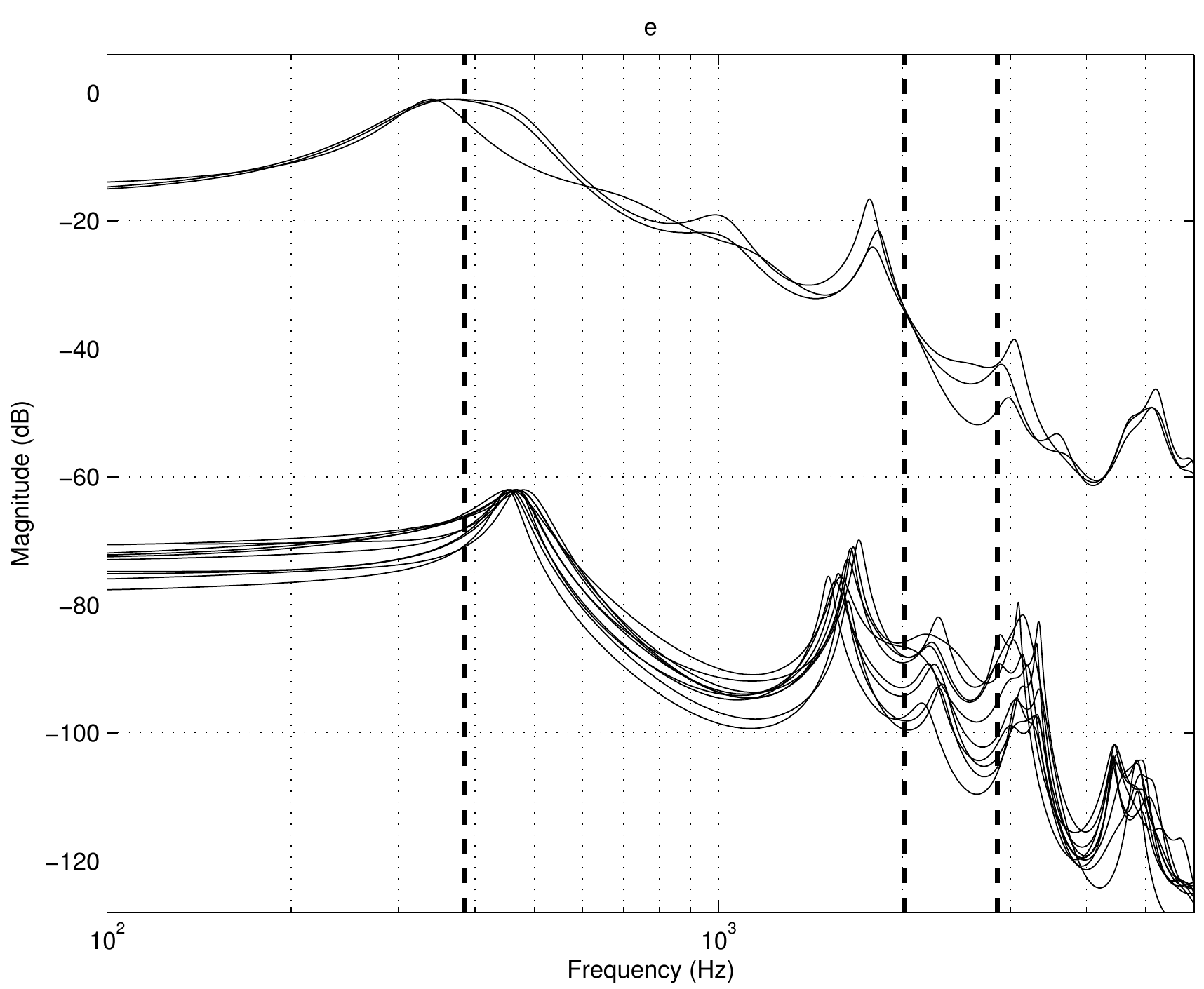}
\caption{Spaghetti plots of LPC envelopes from Finnish front vowels
  [\textipa{i, y, u, e}] in the order of increasing $F_1$.  In each
  panel, the upper graphs have been produced from recordings during
  the MRI using the noise reduction detailed in
  Section~\ref{AlgorithmSec}. The lower graphs have been produced from
  recordings in anechoic chamber from the same test subject.  The
  vertical dotted lines are the three lowest resonance frequencies
  $R_1, R_2$, and $R_3$, computed by FEM from one of the MRI
  geometries using the Helmholtz model in \cite{H-L-M-P:WFWE}.}
\label{SpectralEnvelopesFront}
\end{figure}

A number of LPC envelopes, produced by the MATLAB function \verb lpc ,
for each the eight Finnish vowel [\textipa{\textscripta , e, i, o, u,
    y, \ae , \oe}] are given in
Figs.~\ref{SpectralEnvelopesFront}--\ref{SpectralEnvelopesBack}. All
the data has been recorded from a healthy 26-year-old male in supine
position. There are samples during an MRI scan as well as comparison
samples that have been recorded in an anechoic chamber.  Three lowest
acoustic resonances have been computed by FEM, based on the vocal
tract geometries obtained by MRI.

\subsubsection*{Sound data during MRI}

As pointed out in Section~\ref{IntroSec}, a second set of pilot MRI
experiments was carried out in 2012. The test subject was able to
produce $107$ speech samples during a single MRI session of $1.5$ h
according to the experimental specifications given in
Section~\ref{PhonExpSec}. Out of these speech and MRI samples, $69$
are vowels
imaged by static 3D MRI, out of which $40$ with $f_0 \approx 104$~Hz
were chosen for the validation experiment.
 
The vowel samples were processed by the noise reduction algorithm
detailed in Section~\ref{AlgorithmSec}, and their LPC envelopes (using
filter order $40$) shown in
Figs.~\ref{SpectralEnvelopesFront}--\ref{SpectralEnvelopesBack} were
produced by MATLAB. In many but not in all cases, the lowest formants
$F_1, F_2,$ and $F_3$ could be correctly revealed by Praat
\cite{Praat:2010} (using default settings) from the de-noised signals.

\subsubsection*{Comparison sound data from anechoic chamber}

To obtain high-quality comparison data, speech samples were recorded
in anechoic chamber from the same subject in supine position.  Brüel
\& Kj\ae{}ll 2238 Mediator integrating sound level meter was used as a
microphone, coupled to RME Babyface digitizer with software
TotalMix~FX~v.0.989 and Audacity~v.1.3.14 running on
MacBook~Air~OSX~10.7.5. The test subject heard from earphones his own,
algorithmically de-noised speech signal from pilot MRI experiments as
a pitch reference. The vowels were given in a randomised order and
also shown on a computer screen.

Even though these experiments were designed to resemble the conditions
during the MRI scan in many respects, there are significant
differences. Firstly, the acoustic noise of the MRI machine was not
replicated in the anechoic chamber. Secondly, the test subject fatigue
played lesser role in anechoic chamber since the total duration of a
single experimental session was only about 10 minutes. Thirdly, the head
and neck MRI coil is a rather closed acoustic environment whereas
there was no similar acoustic load present in the anechoic
chamber. Reflections from the MRI coil walls produce spurious
``external formants'' to measured speech signals, and we believe this
to be the explanation for some of the extra peaks that can be seen the
upper curves in
Figs.~\ref{SpectralEnvelopesFront}--\ref{SpectralEnvelopesBack}.

\subsubsection*{Computation of the Helmholtz resonances}
 
\begin{figure}[t]
\center
\includegraphics[width=55mm,height=40mm]{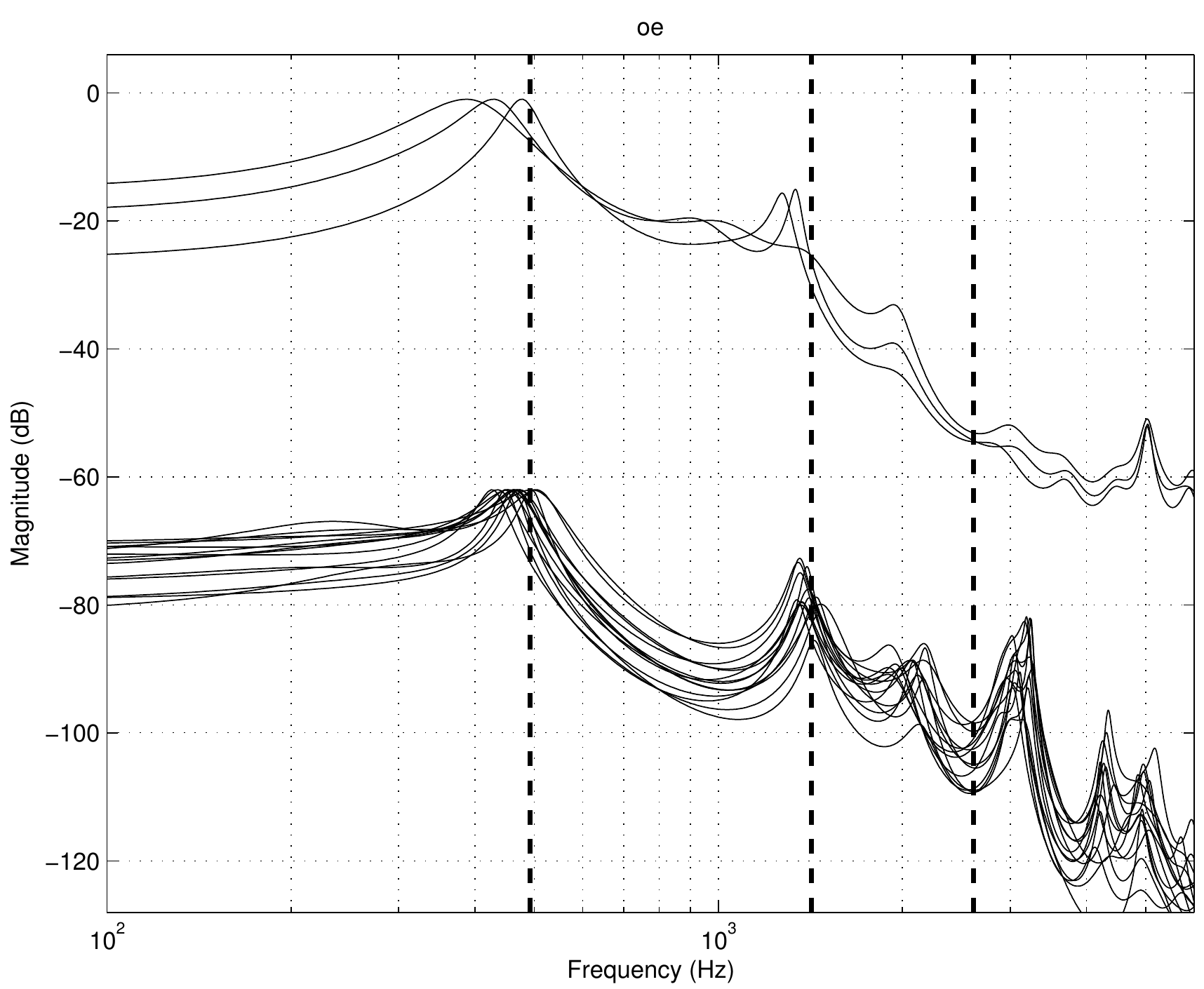}
\includegraphics[width=55mm,height=40mm]{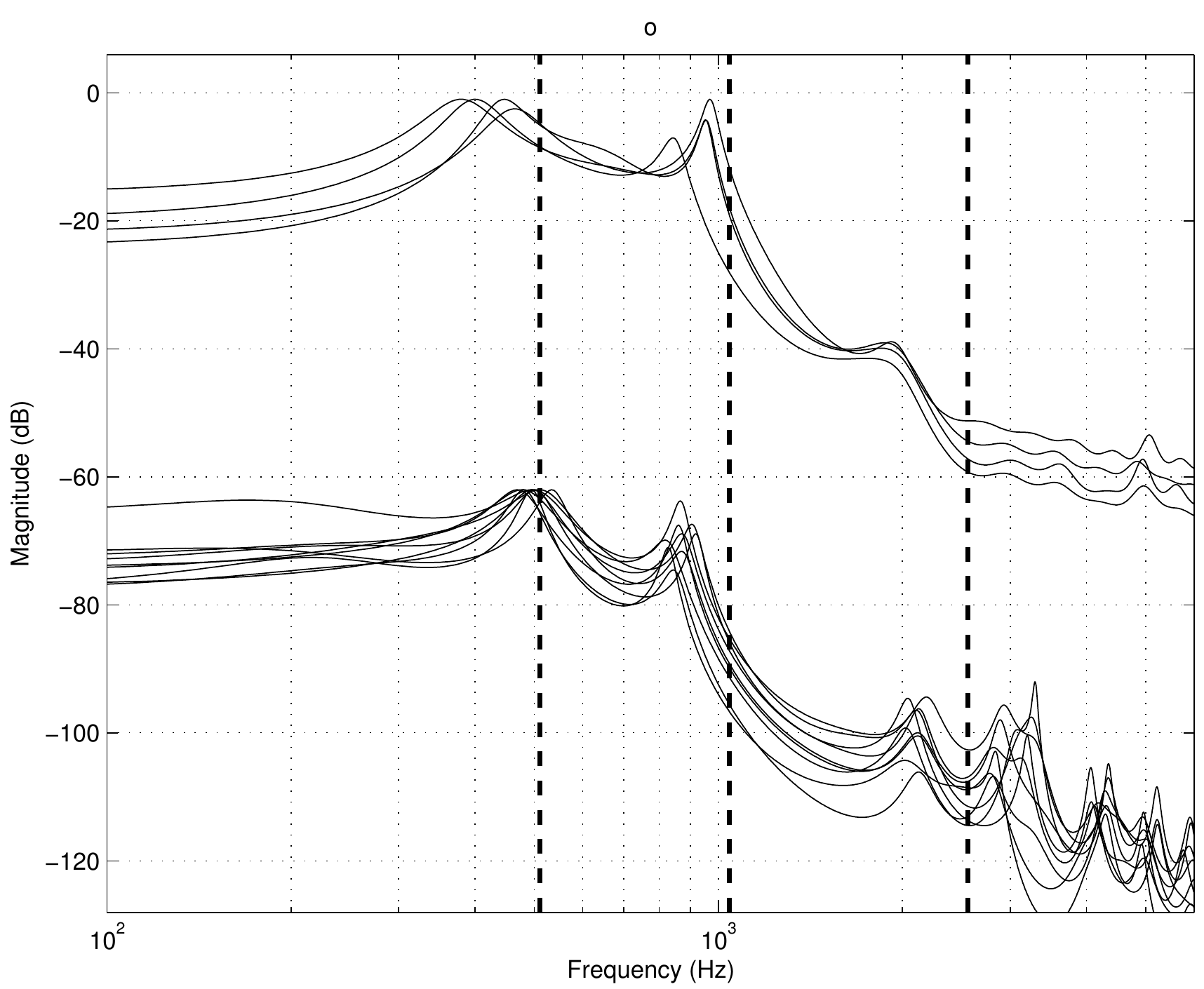}
\center
\includegraphics[width=55mm,height=40mm]{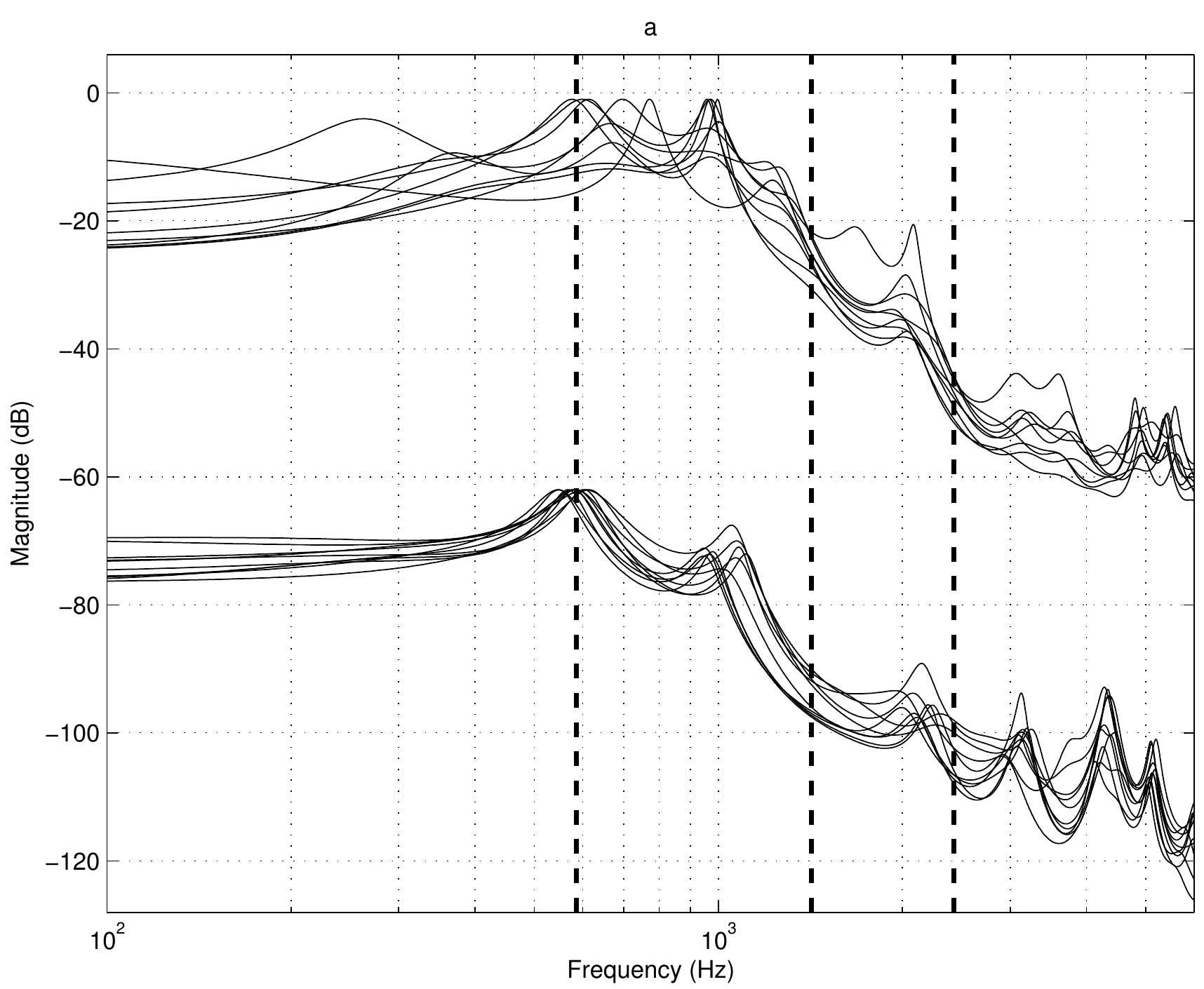}
\includegraphics[width=55mm,height=40mm]{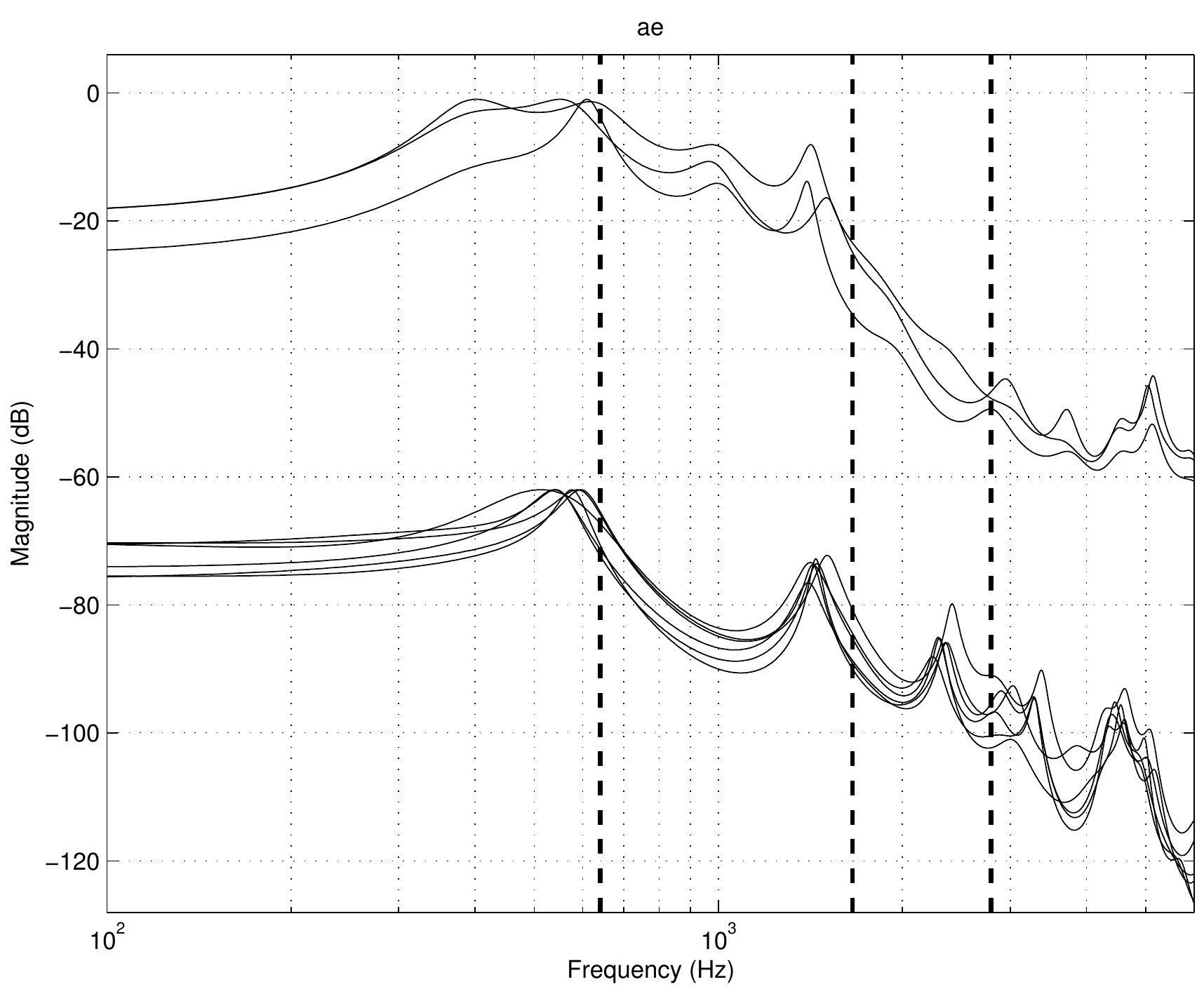}
\caption{Spaghetti plots of LPC envelopes of  Finnish back vowels
  [\textipa{\oe, o, \textscripta , \ae }].  The
  presentation is similar with Fig.~\ref{SpectralEnvelopesFront}.}
\label{SpectralEnvelopesBack}
\end{figure}

For each vowel presented in
Figs.~\ref{SpectralEnvelopesFront}--\ref{SpectralEnvelopesBack}, one
MRI scan was randomly chosen from the full set. This MRI geometry was
processed as described in \cite{A-H-H-K-M-S-R:ASEMRIDMHVT} to produce
surface models of the air-tissue interface as shown in
Fig.~\ref{JuhaGeom}. The models did not include teeth geometries at
all.

The FEM mesh was generated from the surface model, and the acoustic
resonances $R_1, R_2 \ldots$, of the vocal tract air column were
computed using the Helmholtz resonance model detailed in
\cite{H-L-M-P:WFWE}. The Dirichlet boundary condition (which is a very
rudimentary exterior space acoustic model indeed) was used at the
mouth, leading to an overestimation of $F_2$ and $F_3$ by the
respective $R_2$ and $R_3$. The FEM computation reveals a cloud of
higher Helmholtz resonances $R_4, R_5, R_6, \ldots$ near the expected
fourth formant $F_4$ position (as given by Praat~v.4.6.15 or
\verb|lpc| in MATLAB) but they are not shown in
Figs.~\ref{SpectralEnvelopesFront}--\ref{SpectralEnvelopesBack}.

\section{\label{ConclSec}Conclusions}

We have described experimental protocols, MRI sequences, a sound
recording system, and a customised post-processing algorithm for
contaminated speech that, in conjunction with previously reported
arrangements
~\cite{A-H-M-P-P-S-V:RSSAMRI,L-M-P:RSDMRI,M-P:RSDMRIII,Palo:WEM:2011},
can be used for simultaneous speech sound and anatomical data
acquisition on a large number of oral and maxillofacial surgery
patients.

The data set obtained from these measurements are primarily intended
for parameter estimation, fine tuning, and validation of a
mathematical model for speech production. However, these methods and
procedures may be used in a wider range of applications, including
medical research and clinical use.

Collecting such multi-modal data from a large set of patients is far
from a trivial task even when suitable instrumentation is
available. Several phonetic aspects must be taken into account to
ensure that the task is within the ability of the patients, regardless
of background and skills. It must be possible to monitor the quality
of articulation and phonation despite the acoustic noise in the MRI
room, and data collection procedures must be reliable to minimise the
number of repetitions and the amount of useless data obtained.  All
this must be achieved in as short a time as possible to minimise cost
and maintain patient interest in the project.

The experimental setting and phonetic tasks require the patients to
have abilities in consentration, remaining still, and sustaining
prolonged phonation not significantly reduced from young adults in
good health. At the time of writing of this article, seven patients
(out of which four are female) have already undergone such MRI
examinations preceding their orthognatic procedures, and they are
expected to take part in a similar examination after their
post-operative treatment will have been completed. All of these
examinations have succeeded without any kind of complications, and the
resulting MRI image and the speech sound data quality is very
satisfactory as well. Applications to other patient groups are under
consideration but may require adaptations to the required time of
phonation and the total number of measurements.

Some questions and problems in the measurement arrangements remain
open, in particular, involving acoustic noise and its impact on
articulation. Acoustic noise during measurements remains a problem
from two points of view. Firstly, formant extraction from de-noised,
prolonged vowel samples is sometimes problematic as observed in
Section~\ref{FormantSec}. On the other hand, reliable formant
extraction may be difficult for reasons unrelated with noise
contamination: consider, e.g., vowels with low $F_1$ in high pitch
speech samples such as [\textipa{i}] pronounced by female subjects.
Secondly, the onset of MRI noise may cause a significant adaptation in
the patients' articulation. It may be possible to reduce this problem
by running the 3D MRI sequence once while the patient receives the
task instructions to adapt the patient to the noise, and a second time
during phonation to obtain the vocal tract geometry. For the 2D
sequences, the sequence may be started before phonation for the same
effect.

\section{Acknowledgements}
The authors were supported by the Finnish Academy grant Lastu 135005,
the Finnish Academy projects 128204 and 125940, European Union grant
Simple4All, Aalto Starting Grant, Magnus Ehrnrooth Foundation, {\AA}bo
Akademi Institute of Mathematics, and Research Grant of Turku
University Hospital.  

The authors wish to thank professor P.~Alku, Department of Signal
Processing and Acoustics, Aalto University, for providing resources
and consultation for the acquisition of comparison data in
Figs.~\ref{SpectralEnvelopesFront}--\ref{SpectralEnvelopesBack}.

The current version of the post-processing software described in this
article can be obtained from the authors by request. The raw MRI data,
the speech sound data, and the surface models related to
Figs.~\ref{SpectralEnvelopesFront}--\ref{SpectralEnvelopesBack} are
freely available for non-commercial academic and educational use.

\bibliographystyle{plain}


\begin{thebibliography}{10}

\bibitem{Aalto:dt:2009}
A.~Aalto.
\newblock A low-order glottis model with nonturbulent flow and mechanically
  coupled acoustic load.
\newblock Master's thesis, Aalto University, Department of Mathematics and
  Systems Analysis, 2009.

\bibitem{A-A-M-M-V:MLBVFVTOPI}
A.~Aalto, D.~Aalto, J.~Malinen, and M.~Vainio.
\newblock Modal locking between vocal fold and vocal tract oscillations.
\newblock arXiv:1211.4788 (submitted), 2013.

\bibitem{A-A-M:LFPSGFM}
A.~Aalto, P.~Alku, and J.~Malinen.
\newblock A {LF}-pulse from a simple glottal flow model.
\newblock In {\em Proceedings of the 6th International Workshop on Models and
  Analysis of Vocal Emissions for Biomedical Applications (MAVEBA2009)}, pages
  199--202, 2009.

\bibitem{A-H-M-P-P-S-V:RSSAMRI}
D.~Aalto, O.~Aaltonen, R.-P. Happonen, J.~Malinen, P.~Palo, R.~Parkkola,
  J.~Saunavaara, and M.~Vainio.
\newblock Recording speech sound and articulation in {MRI}.
\newblock In {\em Proceedings of BIODEVICES 2011}, pages 168--173, 2011.

\bibitem{A-H-H-K-M-S-R:ASEMRIDMHVT}
D.~Aalto, J.~Helle, A.~Huhtala, A.~Kivel\"a, J.~Malinen, J.~Saunavaara, and
  T.~Ronkka.
\newblock Algorithmic surface extraction from {MRI} data: modelling the human
  vocal tract.
\newblock In {\em Proceedings of BIODEVICES 2013}, 2013.

\bibitem{A-H-K-M-P-S-V:HFAVFFCVTR}
D.~Aalto, A.~Huhtala, A.~Kivel\"a, J.~Malinen, P.~Palo, J.~Saunavaara, and
  M.~Vainio.
\newblock How far are vowel formants from computed vocal tract resonances?
\newblock arXiv:1208.5963, 2012.

\bibitem{A-M-V-S-P:EMAEMRID}
D.~Aalto, J.~Malinen, M.~Vainio, J.~Saunavaara, and J.~Palo.
\newblock Estimates for the measurement and articulatory error in {MRI} data
  from sustained vowel phonation.
\newblock In {\em Proceedings of the International Congress of Phonetic
  Sciences}, 2011.

\bibitem{SB:SANSUSS}
S.~Boll.
\newblock Suppression of acoustic noise in speech using spectral subtraction.
\newblock {\em Acoustics, Speech and Signal Processing, IEEE Transactions on},
  27(2):113 -- 120, 1979.

\bibitem{Chiba:VNS:1958}
T.~Chiba and M.~Kajiyama.
\newblock {\em The Vowel, Its Nature and Structure}.
\newblock Phonetic Society of Japan, 1958.

\bibitem{Dedouch:FEM:2002}
K.~Dedouch, J.~Hor\'a\v{c}ek, T.~Vampola, and L.~\v{C}ern\'y.
\newblock Finite element modelling of a male vocal tract with consideration of
  cleft palate.
\newblock In {\em Forum Acusticum}, 2002.

\bibitem{Dunn:CVR:1950}
H.~K. Dunn.
\newblock The calculation of vowel resonances, and an electrical vocal tract.
\newblock {\em J. Acoust. Soc. Am.}, 22:740 -- 753, 1950.

\bibitem{ElMasri:DTL:1998}
S.~El~Masri, X.~Pelorson, P.~Saguet, and P.~Badin.
\newblock Development of the transmission line matrix method in acoustics.
  applications to higher modes in the vocal tract and other complex ducts.
\newblock {\em Int. J. of Numerical Modelling}, 11:133 -- 151, 1998.

\bibitem{Ericsdotter:AAR:2005}
C.~Ericsdotter.
\newblock {\em Articulatory-Acoustic Relationships in Swedish Vowel Sounds}.
\newblock PhD thesis, Stockholm University, Stockholm, Sweden, 2005.

\bibitem{Fant:ATS:1960} G.~Fant.  
\newblock {\em Acoustic Theory of  Speech Production}.  
\newblock Mouton, the Hague, 1960.

\bibitem{H-L-M-P:WFWE}
A.~Hannukainen, T.~Lukkari, J.~Malinen, and P.~Palo.
\newblock Vowel formants from the wave equation.
\newblock {\em Journal of the Acoustical Society of America Express Letters},
  122(1):EL1--EL7, 2007.

\bibitem{Helmholtz:LT:1863}
H.~L.~F. Helmholtz.
\newblock {\em Die Lehre von den Tonempfindungen als physiologische Grundlage
  für die Theorie der Musik}.
\newblock Braunschweig: F. Vieweg, 1863.

\bibitem{H-U-R-V-B:AVMHGNSOVFM}
J.~Hor\'a\v{c}ek, V.~Uruba, V.~Radolf, J.~Vesel\'y, and V.~Bula.
\newblock Airflow visualization in a model of human glottis near the
  self-oscillating vocal folds model.
\newblock {\em Applied and Computational Mechanics}, 5:21--28, 2011.

\bibitem{Playrec}
R.~Humphrey.
\newblock http://www.playrec.co.uk/.
\newblock {A}ccessed~Jul.~15th,~2012.

\bibitem{Kelly:SS:1962}
J.L. Kelly and C.C. Lochbaum.
\newblock Speech synthesis.
\newblock In {\em Proceedings of the 4th International Congress on Acoustics},
  Paper G42: 1--4, 1962.

\bibitem{Lu:FES:1993}
C.~Lu, T.~Nakai, and H.~Suzuki.
\newblock Finite element simulation of sound transmission in vocal tract.
\newblock {\em J. Acoust. Soc. Jpn. (E)}, 92:2577 -- 2585, 1993.

\bibitem{L-M-P:RSDMRI}
T.~Lukkari, J.~Malinen, and P.~Palo.
\newblock Recording speech during magnetic resonance imaging.
\newblock In {\em Proceedings of the 5th International Workshop on Models and
  Analysis of Vocal Emissions for Biomedical Applications (MAVEBA2007)}, pages
  163--166, 2007.

\bibitem{JM:LPATR}
J.~Makhoul.
\newblock Linear prediction: A tutorial review.
\newblock {\em Proceedings of the {IEEE}}, 63(4):561--580, 1975.

\bibitem{Makhoul:SLP:1975}
J.~Makhoul.
\newblock Spectral linear prediction: Properties and applications.
\newblock {\em IEEE Transactions on Acoustics, Speech and Signal Processing},
  23(3):283 -- 296, 1975.

\bibitem{M-P:RSDMRIII}
J.~Malinen and P.~Palo.
\newblock Recording speech during {MRI}: {P}art {II}.
\newblock In {\em Proceedings of the 6th International Workshop on Models and
  Analysis of Vocal Emissions for Biomedical Applications (MAVEBA2009)}, pages
  211--214, 2009.

\bibitem{Mullen:WPM:2006}
J.~Mullen, D.~Howard, and D.~Murphy.
\newblock Waveguide physical modeling of vocal tract acoustics: Flexible
  formant bandwith control from increased model dimensionality.
\newblock {\em IEEE Transactions on Audio, Speech and Language Processing},
  14(3):964 -- 971, 2006.

\bibitem{Niemi:AC:2006}
M.~Niemi, J.P.~Laaksonen, T.~Peltomaki, J.~Kurimo, O.~Aaltonen, and RP~Happonen.
\newblock Acoustic comparison of vowel sounds produced before and after
  orthognathic surgery for mandibular advancement.
\newblock {\em Journal of Oral \& Maxillofacial Surgery}, 64(6):910--916, 2006.

\bibitem{Nishimoto:ETF:2004}
H.~Nishimoto, M.~Akagi, T.~Kitamura, and N.~Suzuki.
\newblock Estimation of transfer function of vocal tract extracted from {MRI}
  data by {FEM}.
\newblock In {\em The 18th International Congress on Acoustics}, volume~II,
  pages 1473--1476, 2004.

\bibitem{Opto:2009}
{Optoacoustics,~Ltd.}
\newblock http://www.optoacoustics.com/, 2009.
\newblock {A}ccessed~Aug.~25th,~2009.

\bibitem{P-A-A-H-M-S-V:AFVRMRISD}
J.~Palo, D.~Aalto, O.~Aaltonen, R.-P. Happonen, J.~Malinen, J.~Saunavaara, and
  M.~Vainio.
\newblock Articulating {F}innish vowels: Results from {MRI} and sound data.
\newblock {\em Linguistica Uralica}, 48(3):194--199, 2012.

\bibitem{Palo:WEM:2011}
P.~Palo.
\newblock {\em A wave equation model for vowels: Measurements for validation}.
\newblock Licentiate thesis, Aalto University, Institute of Mathematics, 2011.

\bibitem{Praat:2010}
{Praat v.4.6.15}.
\newblock http://www.fon.hum.uva.nl/praat/.
\newblock {A}ccessed~July.~5th,~(\textbf{2010}).

\bibitem{P-H-H:TMMNRRSDPMRID}
J.~P\v{r}ibil, J.~Hor\'a\v{c}ek, and P.~Hor\'ak.
\newblock Two methods of mechanical noise reduction of recorded speech during
  phonation in an {MRI} device.
\newblock {\em Measurement science review}, 11(3), 2011.

\bibitem{P-P-F:ASPANPDMRI}
J.~P\v{r}ibil, A.~P\v{r}ibilov\'a, and I.~Frollo.
\newblock Analysis of spectral properties of acoustic noise produced during
  magnetic resonance imaging.
\newblock {\em Applied Acoustics}, 73(8):687--697, 2012.

\bibitem{Rofsky:VIBE:1999}
N.~Rofsky, V.~Lee, G.~Laub, M.~Pollack, G.~Krinsky, D.~Thomasson, M.~Ambrosino, and J.~Weinreb.
\newblock Abdominal {MR} imaging with a volumetric interpolated breath-hold
  examination.
\newblock {\em Radiology}, 212(3):876 -- 884, 1999.

\bibitem{Suzuki:SVT:1993}
H.~Suzuki, T.~Nakai, N.~Takahashi, and A.~Ishida.
\newblock Simulation of vocal tract with three-dimensional finite element
  method.
\newblock {\em Tech. Rep. IEICE}, EA93-8:17 -- 24, 1993.

\bibitem{Vahatalo:EG:2005}
K.~Vahatalo, J.P.~Laaksonen, H.~Tamminen, O.~Aaltonen, and RP~Happonen.
\newblock Effects of genioglossal muscle advancement on speech: an acoustic
  study of vowel sounds.
\newblock {\em Otolaryngology - Head \& Neck Surgery}, 132(4):636--640, 2005.

\bibitem{V-S-J-J-M:DSITBMSRTNEF}
M.~Vainio, A.~Suni, H.~J{\"a}rvel{\"a}inen, J.~J{\"a}rvikivi, and V.~Mattila.
\newblock Developing a speech intelligibility test based on measuring speech
  reception thresholds in noise for {E}nglish and {F}innish.
\newblock {\em J. Acoust. Soc. Am.}, 118:1742--1750, 2005.

\bibitem{S-H-R:PCFDSF3DM}
P.~\v{S}idlof, J.~Hor\'a\v{c}ek, and V.~\v{R}idk\'y.
\newblock Parallel {CFD} simulation of flow in a {3D} model of vibrating human
  vocal folds.
\newblock {\em Computers and Fluids}, 2012.

\bibitem{Svancara:NME:2006}
P.~\v{S}vancara and J.~Hor\'a\v{c}ek.
\newblock Numerical modelling of effect of tonsillectomy on production of
  {C}zech vowels.
\newblock {\em Acta Acustica united with Acustica}, 92:681 -- 688, 2006.

\bibitem{Svancara:NMP:2004}
P.~\v{S}vancara, J.~Hor\'a\v{c}ek, and L.~Pe\v{s}ek.
\newblock Numerical modelling of production of {C}zech wovel /a/ based on {FE}
  model of the vocal tract.
\newblock In {\em Proceedings of International Conference on Voice Physiology
  and Biomechanics}, 2004.

\end{thebibliography}

\end{document}